\documentclass[aps,prl,showpacs,superscriptaddress,amssymb,nofootinbib,twocolumn]{revtex4}

\usepackage{graphicx}
\usepackage{braket}

\begin{document}

\title{Quantum walk of a trapped ion in phase space}

\author{H.\ Schmitz, R.\ Matjeschk, Ch.\ Schneider, J.\ Glueckert,
M.\ Enderlein, T.\ Huber and T.\ Schaetz}
\email{tobias.schaetz@mpq.mpg.de}
\affiliation{Max-Planck-Institut f\"{u}r Quantenoptik,
Hans-Kopfermann-Strasse 1, D-85748 Garching, Germany}

\begin{abstract}
We implement the proof of principle for the quantum walk of one ion
in a linear ion trap.\ With a single-step fidelity exceeding 0.99,
we perform three steps of an asymmetric walk on the line.\ We
clearly reveal the differences to its classical counterpart if we
allow the walker/ion to take all classical paths simultaneously.\
Quantum interferences enforce asymmetric, non-classical
distributions in the highly entangled degrees of freedom (of coin
and position states).\ We theoretically study and experimentally
observe the limitation in the number of steps of our approach, that
is imposed by motional squeezing.\ We propose an altered protocol
based on methods of impulsive steps to overcome these restrictions,
in principal allowing to scale the quantum walk to several hundreds
of steps.
\end{abstract}


\maketitle
A quantum walk\cite{aharonov93} is the deterministic quantum
mechanical extension of a classical random walk.\ A simple classical
version requires two basic operations: Tossing the coin
(coin-operation), allowing for two possible and random outcomes.\
Dependent on this outcome, the walker performs a step to the right
or left (step-operation).\ In the quantum mechanical extension the
operations allow for coherent superpositions of entangled coin and
position states.\ After several iterations the probability to be in
a certain position is determined by quantum mechanical interference
of the walker state that leads to fundamentally different
characteristics of the walk\cite{kempe03}.\

The motivation for studying quantum walks is twofold.\ On the one
hand, many classical algorithms include random walks.\ Examples can
be found in biology, psychology, economics and physics, for example
Einstein's simple model for Brownian motion\cite{einstein05}.\ The
extension of the walk to quantum mechanics might allow for
substantial speed-up of related quantum versions\cite{kempe03}, as
in prominent algorithms suggested by Shor\cite{shor94} and
Grover\cite{grover96} due to other quantum-subroutines.\ On the
other hand, the quantum walk could lead to new insights into
entanglement and decoherence in mesoscopic systems\cite{monroe96}.\
These topics might be explored by increasing the amount of walkers -
even before any algorithm might benefit from the quantum random
walk.\

Quantum walks have been thoroughly investigated theoretically and
first attempts at implementation have been performed with a very
limited amount of steps due to a lack of operation fidelity or
fundamental restrictions within the protocol.\ Some aspects have
been realized on the longitudinal modes of a linear optical
resonator\cite{bouwmeester99} and in a nuclear magnetic resonance
experiment\cite{laflamme05}.\ An implementation based on neutral
atoms in a spin-dependent optical
lattice\cite{duerr02,mandel03,meschede09} has resulted in an
experiment recently.\ Other proposals considered an array of
microtraps illuminated by a set of microlenses\cite{birkl05} and
Bose-Einstein condensates\cite{chandra06}.\
\begin{figure}
\begin{center}
\includegraphics*[width=\columnwidth]{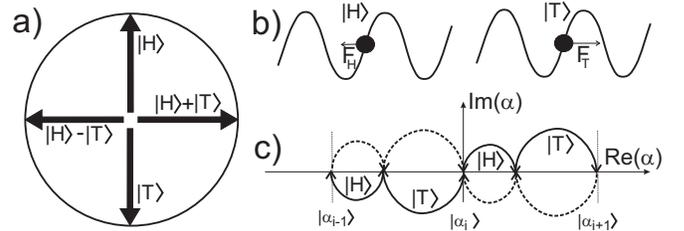}
\end{center}
\vspace*{-0.5cm} \caption{Schematic of the required states and
operations to implement the quantum walk of an ion.\ a) Cross
section of the Bloch sphere of two long lived electronic states
($\ket{H}, \ket{T}$) of the ion, spanning the coin space ($Head$ and
$Tail$). To implement the deterministic ``tossing of the coin'', we
couple the two states via radio-frequency radiation (i.e.\ we rotate
the state vector), additionally introducing the superposition states
($\ket{H}+\ket{T}$) and ($\ket{H} - \ket{T}$) (normalization
omitted).\ b) The coin state dependent optical dipole force
($F_H,F_T$) is modulated sinusoidally and displaces the walker/ion
for state $\ket{H}$ ($\ket{T}$) to the left (right).\ c) The
off-resonantly oscillating forces allow a coherent displacement
along a circular path in the co-rotating phase space.\ The state
$(\ket{H}+\ket{T})\ket{\alpha_i}$ is exposed to two displacement
pulses, each followed by an $\hat{R}(\pi,0)$-pulse to exchange the
populations of states $\ket{H}$ and $\ket{T}$. Despite different
amplitudes of the forces we achieve state
($\ket{H}\ket{\alpha_{i+1}} + \ket{T}\ket{\alpha_{i-1}}$), the
walker/ion stepping to both sides simultaneously with equal step
size. By subsequent coin- (a) and step-operations (b), we implement
the quantum walk along a line.} \label{Figure:1}
\end{figure}
Travaglione and Milburn\cite{milburn02} proposed a scheme for
trapped ions to transfer the high operational
fidelities\cite{monroe96} obtained in quantum information processing
(QIP) into the field of quantum walks.\ We implement the proof of
principle for a discrete, asymmetric quantum walk of one trapped ion
along a line in phase space.\ After three steps, each performed with
a fidelity exceeding 0.99, we reveal the differences to a classical
random walk.\
The limit of coherent displacements to states inside
the Lamb-Dicke regime was foreseen by\cite{milburn02},
experimentally observed
in a different context\cite{stean08}, and is confirmed by us.\\

In the experimental realization, we confine one $^{25}$Mg$^+$ ion of
mass $m$ in a linear multi-zone Paul trap\cite{trap09}.
Motion in the axial (\emph{z}-) direction is harmonic with an
oscillation frequency $\omega_z = 2\pi \cdot 2.1$ MHz and energy
eigenstates $ \ket{n}$.\ The ion's/walker's state $\Psi_N$ after $N$
steps can be described by the state of the coin, $\ket{T}$ and
$\ket{H}$, and positions $\ket{\alpha_i}$ (with $ i $ integer) on
the line, as depicted in Fig.\ 1.\ The coin states are composed of
the $^2S_{1/2}$-ground-state hyperfine levels of the ion, $\ket{
F=3,M_F=3}$ and $\ket{F=2,M_F=2}$, labeled $\ket{T}$ and $\ket{H}$,
respectively (splitting $\omega_0 \simeq 2\pi \cdot 1.8$ GHz, a
magnetic field of $6\cdot 10^{-4}$ Tesla lifts the degeneracy of
states of identical $F$).\
Ideally, the position of the walker is encoded into the coherent
state
\begin{equation}
\vspace{-0.2cm} \ket{ \alpha_i} = e^{-\mid \alpha_i \mid^2/2}
\sum_{n=0}^\infty \frac{\alpha_{i}^n}{\sqrt{n!}} \ket{n},
\label{coherentstate}
\end{equation}
whith  $i$ running from $-N$ to $N$.\

The initial state of each walk - $\Psi_{0}= \ket{T} \otimes \ket{0}$
- is prepared by laser cooling to $\overline{n}(N=0) < 0.03$, close
to the ground state of motion\cite{monroecool95} and optically
pumping into the electronic state $\ket{T}$~\cite{monroecool95}. To
implement the quantum coin toss, we couple state $\ket{T}$ and
$\ket{ H}$ with a resonant radio-frequency(rf) pulse.\ The pulse
implements a rotation of the coin-state on the Bloch
sphere\cite{eberly87}
\begin{eqnarray}
\hat{R}(\Theta, \Phi)= \left(
\begin{array}{cc}
  \mathrm{cos}(\small{\frac{\Theta}{2}}) & -i e^{-i \Phi}\mathrm{sin}(\small{\frac{\Theta}{2}}) \\
  -i e^{+i \Phi}\mathrm{sin}(\small{\frac{\Theta}{2}})& \mathrm{cos}(\small{\frac{\Theta}{2}}) \\
\end{array}%
\right),
\end{eqnarray}
where $\ket{T} = (0,1)^T$ and $\ket{H} = (1,0)^T$.\ The angle
$\Theta$ is proportional to the duration of the rf-pulse and $\Phi$
is its phase.\ For the coin-operation we chose
$\hat{R}(\pi/2,-\pi/2)$ with the rotation axis perpendicular to the
cross-section of the Bloch sphere depicted in Fig.\ 1a.

The conditional unitary step-operator
\begin{equation}
\vspace{-0.2cm} \hat{S} \! = \! \sum_{i} (\ket{T}  \bra{T}
\small{\otimes} \ket{ \alpha_{i+1}} \bra{ \alpha_{i}} + \ket{H}
\bra{H} \small{\otimes} \ket{ \alpha_{i-1}} \bra{ \alpha_{i}})
\label{stepop}
\end{equation}
is realized by a state dependent optical dipole
force\cite{monroe96,bible98,london03}.\ For this, two Raman laser
beams at frequencies $\omega_1$ and $\omega_2$, both detuned
$(2\pi\cdot 80\,$GHz) blue from the $^2S_{1/2}\rightarrow \!
^{2}P_{3/2}$ electric dipole transition ($\lambda \approx 280\!$ nm)
induce an AC-Stark shift of the electronic states.\ Their
lin$\perp$lin polarization and directions are chosen to provide
forces $(F_T=-\frac{3}{2} F_H)$ on the walker/ion that are
conditioned on the coin state, as depicted in Fig.\ 1b.
\begin{figure}
\begin{center}
\includegraphics*[width=\columnwidth]{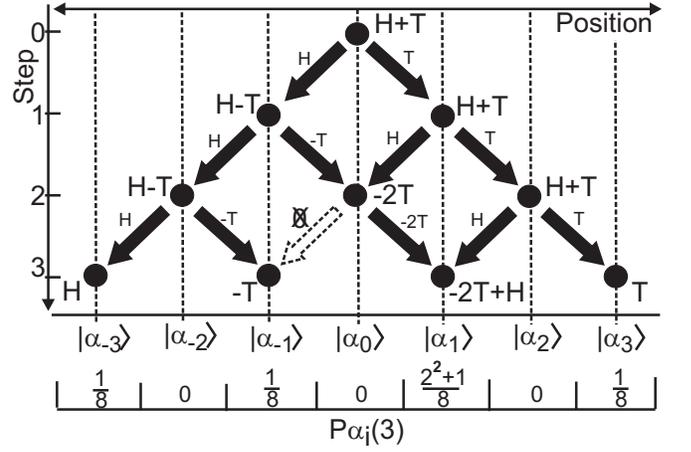}
\end{center}
\vspace*{-0.5cm} \caption{Schematic illustrating the protocol of the
asymmetric walk for three steps.\ With the coin in a superposition
state ($\ket{H}+\ket{T}$) and the position in phase space in the
motional ground state $\ket{ \alpha_0} \equiv \ket{ n=0}$, we
perform the first step.\ We conditionally displace the walker/ion
and therefore encode the superposition of the coin state in a
superposition of positions ($\ket{\alpha_{\pm 1}}$) in phase space.\
Before each step we apply the coin-operation, again providing
superpositions of the coin states of different phases and the
step-operation, allowing for interference between all possible
paths.\ During step $\#$3, interference effects lead to the
asymmetric\cite{asym} probability distribution of position
($P_{\alpha_{i}}(N=3)$) and coin states
 ($P_{T}(N=3)=|\braket{ T |
\Psi_{3}}|^2 = (1+2^2+1)/8$, $P_{H}(N=3)=|\braket{H|\Psi_{3}}|^2 =
(1+1)/8$).
 Their values clearly differ from the
classical symmetric equivalent.} \label{Figure:2}
\end{figure}
If $\omega_2 - \omega_1 = \omega_z$, the forces would resonantly
drive the ion, leading to coherent displacements $\ket{ \alpha_i}$
along a line in phase space (co-rotating at $\omega_z$ within this
letter).\ In our implementation, we choose a detuning $\delta =
\omega_2 - \omega_1 - \omega_z \approx 2 \pi \cdot 100\!$ kHz that
the forces initially dephase and then rephase after a duration
$t_d=2 \pi/ \delta \approx 10\!$\ $\mu s$.\ We stop the coherent
drive after $t_d/2$, then do a $\hat{R}(\pi,0)$-pulse to exchange
$\ket{T}$ and $\ket{H}$ and then drive for $t_d/2$ again and finish
the step sequence with another $\hat{R}(\pi,0)$-pulse. The
$\hat{R}(\pi,0)$-pulses also serve as spin-echos\cite{bible98} that
make our implementation less susceptible to coin state dephasing,
e.g.\ due to magnetic field fluctuations.\ Ideally, the walk of
$N$-steps could be performed by iterating coin- and step-operations
$N$-times, with the ideal evolution depicted in Fig.\ 2.

After finishing a walk sequence, we detect the coin state by
(electronic-) state-dependent fluorescence
measurements\cite{london03}.\ A closed cycling transition on
$^2S_{1/2}\rightarrow$ $^{2}P_{3/2}$ driven by a laser beam on
resonance provides a fluorescence rate of 200\,kHz for state
$\ket{T}$ and negligible fluorescence for state $\ket{H}$
(off-resonant by $\omega_0$). Averaging over many experiments we
determine the probabilities $P_{T}^e(N)$ and $P_{H}^e(N)$ ($^{e}$
indicating an experimental result).\

To access the motional state populations $P_{\alpha_i}^e(N)$ for the
part of the wave function related to $\ket{T}$,
$\braket{T|\Psi^e_N}$, we transfer the population of state $\ket{H}$
by two subsequent rf-pulses similar to the operation described in
Equ.\ 2 via state $^2S_{1/2} \ket{F = 2,M_F =1}$ to $^2S_{1/2}
\ket{F=3,M_F =0}$ without affecting the motional state
distributions.\ To determine $P_{\alpha_i}^e(N)$ associated with
coin state $\ket{H}$, a resonant rf-pulse $\hat{R}(\pi,0)$ is
applied to exchange $\ket{H}\leftrightarrow \ket{T}$ before the
transfer pulses.\

To determine $P_{\alpha_i}^e(N)$ in each case we detune the Raman
beams to the first blue sideband $\omega_1-\omega_2=\omega_0 +
\omega_z $ and drive a two-photon stimulated Raman transition
$\ket{T,n}\leftrightarrow \ket{H,n+1}$.\ In the Lamb-Dicke regime
($\eta^2 n \ll 1$, where the Lamb-Dicke parameter is defined by
$\eta \equiv \Delta k z_0$, with $\Delta k$ is the modulus of the
wave-vector difference of the two Raman beams along $z$, and $z_0 =
\sqrt{\hbar / (2m \omega_z)}$ is the spread of the ground state of
the motional wave function $\ket{\alpha_0}$\cite{bible98}), the Rabi
frequencies can be written as $\Omega_{n,n+1} = \sqrt{n+1} \eta
\Omega$, where $\Omega$ is the Rabi frequency of the carrier
transition $\ket{T,0}\leftrightarrow \ket{H,0}$.\  Typical values in
our implementation are $\Omega \approx 2 \pi \cdot 500\!$ kHz, $z_0
\approx 10\!$ nm and $\eta =0.31$.\
This experiment is performed $> 10^3$ times for each set of
parameters to measure the average probability $P_T^e(N)$ of
occupation in state $\ket{T}$ and repeated for step wise increased
pulse duration $t$ on the blue-sideband\cite{meekhof96}.\ A discrete
Fourier analysis of the recorded fluorescence data allows to deduce
the relative contributions of specific frequency components
$\Omega_{n,n+1}$ and therefore provides $P_{\alpha_i}^e(N)$.\
Since the frequency separation and therefore resolution is maximal
between states $\ket{n=0}$ and $\ket{n=1}$, we can increase our
experimental sensitivity by coherently displacing the motional state
to be measured, $\ket{\alpha_i}$, $i$-times back on the
$\ket{\alpha_0}$ state.
\begin{figure}
\includegraphics*[width= 0.8 \columnwidth]{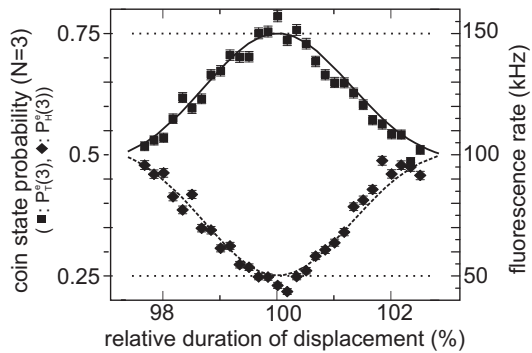}
\vspace*{0 cm} \caption{Probability $P^e_{H}(3)= \vert\braket{H |
\Psi^e_3}\vert ^2$ ($P^e_{T}(3)= \vert\braket{T | \Psi^e_3}\vert
^2$) to observe the walker/ion in coin state $\ket{H}$ ($\ket{T}$)
after step $\#$3 of the quantum walk in dependence on the relative
duration of displacement.\ State-selective detection provides a
fluorescence rate directly proportional to the probability
$P^e_T(3)$.\
We de/increase the duration of the displacement pulses and therefore
shorten/elongate each semi-circular path compared to the case of
ideal duration depicted in Fig.\ 1c.\ A non-ideal duration reduces
the overlap of the final positions associated with $\ket{T}$ and
$\ket{H}$ in phase space. Distinguishable positions provide "which
path" information that destroys the quantum interference of the coin
and reduces the asymmetry to an equal $P_{T}^e(3) \approx P_{H}^e(3)
\approx 0.5$ at relative deviations of $\pm$2\% from the correct
step duration.\ The solid lines depict the prediction of our
theoretical simulation.\ Each data point represents the average of
1200 experiments, with error bars indicating the statistical error.\
For the correct step duration averaged over 60000 experiments, we
achieve $P_H^e(3) = 0.259 \pm 0.001$ and $P_T^e(3) = 0.741 \pm
0.002$.} \label{Figure:3}
\end{figure}

First, we measure the average probabilities $P_{T}^e(N)$ and
$P_{H}^e(N)$ of occupation in state $\ket{T}$ and $\ket{H}$,
respectively.\ After step $\#$1 and $\#$2, these probabilities
$P_{H}^e(1), P_{H}^e(2) (P_{T}^e(1), P_{T}^e(2))$ to observe state
$\ket{H}$ ($\ket{T}$) are 0.5$\pm 0.01$, identical to the classical
evolution.\ Tossing the coin for the third time reveals the coin
asymmetry due to interference in the quantum walk.\ The ideal state
after step $\#$3 is $\Psi_{3}= \ket{H}\left(\ket{\alpha_{-3}} +
\ket{\alpha_{1}}\right) + \ket{T}\left(- \ket{\alpha_{-1}} -2
\ket{\alpha_{1}} + \ket{\alpha_{3}}\right)$.\ We experimentally
observe $P_{T}^e(3)= 0.741 \pm 0.003$ and $P_{H}^e(3)=0.259 \pm
0.001$ (statistical errors only), clearly deviating from the
classical expectation, in good agreement with the ideal ratio
$P_{T}(3)/P_{H}(3) =3$\cite{bias}.\

To investigate the motional coherence within the walk we can either
undo a certain number of steps and measure the probability to return
to the initial state or, more significant, we can slightly vary the
duration of the walker pulses, thus perturbing the overlap of the
final coherent states (positions).\ The results are depicted in
Fig.\ 3.\ Deviations from the correct walker-pulse duration lead to
distinguishable positions for all different paths of the walker.\
Therefore the coin-state interference is completely lost and
$P_T^e(3) \approx P_H^e(3) \approx 0.5$ for pulse length deviations
of $\pm 2 \%$, in good agreement with our theoretical simulations
(solid line in Fig.\ 3).\
\begin{figure}
\includegraphics*[width= \columnwidth]{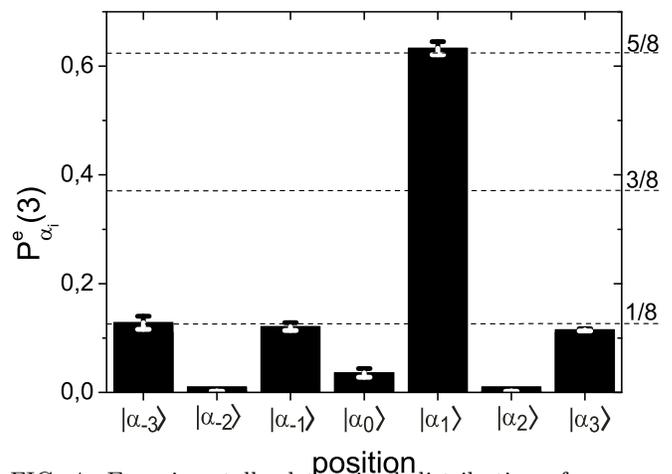}
\vspace*{-1 cm} \caption{Experimentally determined distribution of
occupation probabilities $P_{\alpha_i}^e(3)$ of the positions in
phase space, after step $\#$3 of the random walk.\ Ideally
(classically as well as quantum mechanically), we expect the
populations $P_{\alpha_i}(3)=0$ for even $i$ and $P_{\alpha_{\pm
3}}(3)=1/8$ of the extremal positions each accessible via one out of
8 paths only.\ Classically we expect $P_{\alpha_{\pm 1}}=3/8$.\ For
an optimized step-duration (see Fig.\ 3) we obtain probabilities
very close to the theoretically expected and realistically simulated
values of the quantum walk$, P_{\alpha_{-1}}=1/8$ and
$P_{\alpha_{1}}=5/8$.} \label{Figure:4}
\end{figure}
We further investigate the probabilities $P_{\alpha_i}(N)= |
\braket{\alpha_i|\Psi_N} |^2$ for the positions $\ket{\alpha_i}$ of
the walker/ion.\ After step $\#$1 and $\#$2, we determine
probabilities $P_{\alpha_i}^e(1)$
and $P_{\alpha_i}^e(2)$ identical to the classical ones.\ The
probabilities observed after step $\#$3 are depicted in Fig.\ 4,
with uncertaities due to statistical errors only. As in the
classical counterpart, only one of eight paths reaches each of the
outermost positions and therefore $P_{\alpha_{\pm 3}}(3)=1/8=0.125$.
We find $P_{\alpha_{-3}}^e=0.128 \pm 0.0012$ and
$P_{\alpha_{3}}^e(3)=0.115 \pm 0.002$.\ However, the remaining 6
paths destructively interfere to give $P_{\alpha_{-1}}^e = 0.121 \pm
0.007 $ ($P_{\alpha_{-1}}=1/8$) and constructively add up, leading
to $P_{\alpha_{1}}^e = 0.633 \pm 0.0012$
($P_{\alpha_{1}}=5/8=0.625$).\ The small populations in motional
states of even $i$ result from the residual overlap of the
(non-orthogonal) approximately coherent position-states. This is in
agreement with our theoretical simulations, based on the code of
Ref.\ \cite{stean08}.\ Thus, step $\#$3 again reveals the quantum
nature of the walk.\

In summary, we have implemented the proof of principle for the
quantum walk of one ion with a single step fidelity exceeding 0.99.\
We clearly reveal the differences to a classical counterpart with
experimental uncertainties on the (sub-)percent level.\ The number
of steps is limited by the present protocol.\ More steps lead to
higher motional excitations beyond the Lamb-Dicke regime where
higher order effects in the interaction modify the transition
rates.\ For our actual experimental realization, transition rates
start to decrease around $n=8$ (in our present implementation:
$\overline{n}(1)=1.33;\, \overline{n}(2)=4.71;\,
\overline{n}(3)=9.08$) leading to motional squeezing\cite{meekhof96,
stean08} and state reflection ($\Omega_{n,n+1} \approx 0$) at
$n\approx36$ that cannot be overcome with the implemented walker
steps.\ Reducing $\eta$ by increasing the axial confinement and/or
reducing the angle between the $k$-vectors of the Raman beams, and
thus $\Delta k$, should allow to extend this protocol to $N\leq$ 20
at comparable fidelities.\ Walker states with much higher $N$ can be
achieved by switching to a method different to that
in\cite{monroe96,milburn02}.\
Pairs of resonant, intense and short $\hat{R}(\pi,0)$-pulses will
displace the state of the ion independent of its position in phase
space.\ This would make it possible to walk far outside the
Lamb-Dicke regime with almost arbitrarily many steps, ultimately
limited by the anharmonicity of the trapping potential and/or its
depth, which is of the order of several electron volts.\ However,
read out of position states is not straightforward for large N.\
Incorporating the availability of deterministic entanglement of the
electronic states of more than one ion\cite{didi03,trap09}, more
sophisticated quantum walks become possible, including the "meeting
problem", where two or several entangled particles perform the walk
and experience additional attractive (Bosonic) or repulsive
(Fermionic) interactions\cite{omar06}.
%

%
This work was supported by MPQ, MPG, DFG (Emmy-Noether Programme
Grant No.\ SCHA 973/1-4), SCALA and by the DFG Cluster of Excellence
Munich-Centre for Advanced Photonics.\ We thank Dave Wineland and
Dietrich Leibfried for their invaluable input for the field of
trapped ions in general and for our group, M.J.\ McDonnell for the
very fruitful input on the basics of the simulation code of
Ref.~\cite{stean08} and Ignacio Cirac and Gerhard Rempe for their
great intellectual and financial support.


\begin{thebibliography}{24}
\bibitem{geophase} The area circumscribed in phase space being proportional to an
additional phase\cite{didi03} picked up by the ion is identical for
all pathways and therefore negligible.
\bibitem{bias} An asymmetry of the coin state populations
can also be caused by a low fidelity of the coin state operation.
Anyway, this asymmetry would already arise after step $\#$1 and
$\#$2 and to first order not depend on the duration/lenghth of the
step-operation.
\bibitem{asym} For the symmetric evolution, we can initialize the
walk in the state $\Psi_0 = (\ket{T}+i\ket{H})\ket{\alpha_0}$. Since
our coin-operation $\hat{R}$ does not introduce complex amplitudes,
the paths related to $\ket{T}$ and $\ket{H}$ will not interfere,
providing a symmetric probability distribution.
\end{thebibliography}
\end{document}